\documentclass[12pt]{article}
\usepackage[a4paper,total={18cm,27cm}]{geometry}

\usepackage[utf8]{inputenc}
\usepackage{amsmath}
\usepackage{amsfonts}

\usepackage{graphicx}
\usepackage{array}
\usepackage{caption}

\usepackage{xcolor}
\colorlet{shadecolor}{yellow!20}

\usepackage{authblk}

\usepackage{hyperref}

\usepackage{comment}   % Not used
\newcommand{\NI}{\vspace{0.2cm}\noindent}

\newcommand{\is}{\!=\!}

\begin{document}
%\linenumbers

% **************************************************
% TITLE
% **************************************************

\title{Are cortical microcircuits optimized for information flux ?\\ A simulation-based reverse engineering study}

% **************************************************
% AUTORS
% **************************************************

\author[1]{Claus Metzner}
\author[1]{Ali Ghebleh}
\author[1]{Karin Prebeck}
\author[1,2,3,4]{Achim Schilling}
\author[1]{Andreas Maier}
\author[2]{Thomas Kinfe}
\author[1,2,3,4]{Patrick Krauss}

\affil[1]{\small Cognitive Computational Neuroscience Group, Pattern Recognition Lab, Friedrich-Alexander-University Erlangen-Nürnberg (FAU), Germany}
\affil[2]{\small Neuromodulation and Neuroprosthetics, University Hospital Mannheim, University Heidelberg, Germany}
\affil[3]{\small Neuroscience Lab, University Hospital Erlangen, Germany}
\affil[4]{\small BGU Ludwigshafen, Germany}

\maketitle

% ***************************************************
% ABSTRACT
% ***************************************************

\begin{abstract}
\large
\NI A sufficiently large information flux in recurrent neural networks, quantified by the mutual information between successive network states, is considered a prerequisite for rich information processing capabilities. This raises the question of whether biological neural networks, such as cortical microcolumns, may be structurally organized to enhance information flux. To investigate this possibility, we study a simplified model of the cortical layer 5 architecture, in which a densely and strongly interconnected core population is embedded within a larger supporting network. Surprisingly, we find that the embedding network exerts a pronounced flux-enhancing effect on the core dynamics. Systematic reverse-engineering analyses reveal that the embedding network provides two key contributions: first, it generates effective biases that shift core neurons into a higher-entropy operating regime; second, it supplies stochastic fluctuations that prevent the network from becoming trapped in simple fixed-point or oscillatory attractors through the mechanism of Recurrence Resonance. We further show that the information flux can be increased even beyond the biologically embedded case by applying individually optimized biases to the core neurons, and that these biases can emerge from a simple self-organization principle. Our findings are relevant both for the functional interpretation of biological neural circuits and for the design of artificial recurrent systems such as reservoir computers.

\end{abstract}

% *******************************************************
\newpage

% *******************************************************
\section{Introduction}
% *******************************************************

\NI Over the past decade, deep learning has achieved remarkable advances \cite{lecun2015deep,alzubaidi2021review}, driven in part by the rise of large language models \cite{min2023recent}. Most of these systems are based on feedforward architectures, where signals propagate in a single direction from inputs to outputs. In contrast, recurrent neural networks (RNNs) incorporate feedback loops, allowing them to operate as autonomous dynamical systems \cite{maheswaranathan2019universality}. Once activated, such networks can sustain and transform internal activity even in the absence of ongoing external stimulation.

\NI RNNs possess several fundamental properties that underscore their versatility. They are universal function approximators \cite{schafer2006recurrent} and can emulate arbitrary dynamical systems \cite{aguiar2023universal}. These theoretical capabilities, combined with their practical relevance, have motivated extensive efforts to elucidate their internal working principles. By virtue of their recurrent connectivity, RNNs can accumulate and process information across extended temporal horizons \cite{jaeger2001echo,schuecker2018optimal,busing2010connectivity,dambre2012information,wallace2013randomly,gonon2021fading}. Moreover, they are capable of forming compact yet expressive state representations, shaped by an interplay between dimensional expansion and compression \cite{farrell2022gradient}.

\NI A substantial body of work has addressed the regulation and stabilization of RNN dynamics, particularly under the influence of intrinsic and extrinsic noise sources \cite{rajan2010stimulus,jaeger2014controlling,haviv2019understanding,molgedey1992suppressing,ikemoto2018noise,krauss2019recurrence,bonsel2021control,metzner2022dynamics}. Beyond their engineering applications, RNNs have also been studied as models for neural computation in the brain \cite{barak2017recurrent}. In this context, sparsely connected architectures—resembling biological circuits with limited synaptic connectivity per neuron \cite{song2005highly}—have been shown to support improved memory properties and efficient internal representations \cite{brunel2016cortical,narang2017exploring,gerum2020sparsity,folli2018effect}.

\NI In earlier studies, we initiated a systematic exploration of how structural characteristics shape RNN behavior, starting from elementary three-neuron motifs \cite{krauss2019analysis}. Building on these insights, we demonstrated that the dynamics of large autonomous networks can be effectively characterized and steered by three key parameters: the width of the weight distribution $w$, the connection density $d$, and the excitation–inhibition balance $b$ \cite{krauss2019weight,metzner2022dynamics}.

\NI We subsequently analyzed RNNs within the framework of reservoir computing, focusing on how dynamical regimes and structural constraints influence computational performance. In \cite{metzner2025nonlinear}, we varied both the degree of neuronal nonlinearity and the strength of recurrent interactions to assess their impact on classification tasks with synthetic data. Surprisingly, reservoirs with very weak internal dynamics and only minimal nonlinearities were still able to generate linearly separable representations for the readout, provided that fine-grained high-dimensional features of the state space were exploited. Although strongly chaotic regimes typically impaired performance, peak accuracy was often observed near transitions between chaotic, oscillatory, and fixed-point regimes, lending support to the “edge of chaos” concept.

\NI In \cite{metzner2025organizational}, we further investigated the functional role of biologically inspired structural motifs in RNNs, both in terms of their autonomous dynamics and their reservoir computing performance. The examined features included Dale’s principle, heterogeneous connectivity patches that introduce local modular structure \cite{sporns2016modular,meunier2010modular}, and reciprocal symmetry akin to Hopfield-type networks \cite{hopfield1982neural}. We found that Dale-compliant output structures and heterogeneous weight patches generally enhanced task performance, whereas strong reciprocal symmetry tended to reduce discriminability by driving the network into rapid saturation. These findings suggest that embedding organizational principles into otherwise random connectivity can yield tangible computational benefits.

\newpage

\NI Beyond static structural design, the computational capacity of RNNs can also be modulated dynamically. Counterintuitively, the deliberate injection of external noise into the reservoir may improve performance \cite{bonsel2021control,schilling2022intrinsic,krauss2016stochastic,krauss2019recurrence,schilling2021stochastic,schilling2023predictive,metzner2024recurrence}. In regimes where the network becomes trapped in fixed points or low-period attractors, appropriately tuned stochastic perturbations can re-enable dynamical exploration. This mechanism increases the recurrent information flux \cite{krauss2019recurrence,metzner2024recurrence}, measured as the mutual information between consecutive global network states, which is essential for many computational tasks . As it turned out, the dependence of information flux on noise intensity follows a resonance-like profile, termed “Recurrence Resonance”, and this mechanism therefore requires an external adjustment of the noise amplitude to reach optimal performance.

\NI In our numerical experiments on random signals in recurrent neural networks, we have so far used normally distributed white noise, added independently to the weighted input of each neuron. In a biological setting, such white Gaussian noise could in principle be approximated by the summed spike trains of a large population of statistically independent neurons.

\NI However, it remains unclear whether temporally correlated random signals, possibly with non-Gaussian distributions, might be more effective in preventing RNNs from becoming permanently trapped in strong attractor states. An even more promising scenario is one in which the RNN operates in a feedback loop with a supporting peripheral network that continuously monitors its dynamics and injects adaptive modulatory signals to maintain a computationally favorable regime.

\NI Indeed, the organization of cortical microcircuits, in particular the layer 5 architecture reported by Song et al. \cite{song2005highly}, exhibits features that appear well suited for such a regulatory function. It consists of a densely interconnected excitatory core embedded in a broader population of weakly connected excitatory neurons and inhibitory interneurons (Fig.\ref{Fig_A}(a)).

\NI In this work, we implement a simplified version of this layer 5 architecture, termed the Embedded Core Model (ECM), using probabilistic Boltzmann neurons. As a first exploratory step, we evaluate the effect of the peripheral and interneuron populations on the core in terms of information flux, again quantified as the mutual information between subsequent states. In a second step, we systematically reverse-engineer the ECM in order to identify the essential conditions that enhance information flux within the core.

% *******************************************************
\section{Methods}
% *******************************************************

\subsection{Embedded Core Model}
\label{MethECM}

\NI To investigate the influence of biologically structured connectivity on network dynamics, we construct a recurrent neural network consisting of $N_t = 125$ nodes. The architecture, depicted in Fig.\ref{Fig_A}(a), is designed to capture key statistical features of cortical microcircuits, in particular the coexistence of a small set of strong connections embedded in a large population of weaker ones, as observed experimentally in cortical networks \cite{song2005highly}.

\NI The network comprises three distinct neuron populations: a subset of $N_c = 10$ excitatory 'core' neurons, a larger population of $N_p = 90$ excitatory 'peripheral' neurons, and a group of $N_i = 25$ inhibitory 'interneurons'. The core neurons are densely and strongly interconnected, whereas the remaining excitatory neurons form a sparsely and weakly connected background network. Interneurons do not connect among themselves but project onto both core and peripheral neurons with negative weights, while receiving excitatory input from both populations.

\NI Connections among excitatory neurons are established such that the overall connection density is fixed to $d = 11.6\%$, consistent with experimentally measured connection probabilities in cortical layer 5 microcircuits \cite{song2005highly}. Synaptic weights $W$ for these connections are drawn from a lognormal distribution,
\begin{equation}
p(W) = \frac{1}{W \sqrt{2\pi\sigma^2}} \exp\left(-\frac{(\ln W - \mu)^2}{2\sigma^2}\right),
\end{equation}
with parameters $\mu = -0.702$ and $\sigma = 0.9355$, as reported for synaptic strength distributions in the same study \cite{song2005highly}. To implement the core--periphery structure, the largest weights sampled from this distribution are preferentially assigned to connections within the core subnetwork, while the remaining weights are distributed randomly among all other excitatory connections. This procedure ensures that the core forms a strongly coupled backbone, whereas the surrounding network is dominated by weaker interactions. The connections from inhibitory neurons to excitatory neurons are introduced with the same overall density $d$ and, as a result of numerical optimization, assigned a uniform negative weight $w_i = -5$. The weight matrix of the core network is shown in Fig.\ref{Fig_A}(b), and the full network weight matrix in Fig.\ref{Fig_A}(c). Self-connections are excluded throughout. This construction provides a controlled abstraction of cortical connectivity, reflecting the experimentally observed organization of cortical circuits.

\NI The network dynamics is governed by stochastic Boltzmann neurons, which all update synchronously. Given the total synaptic input $u$ of a neuron, its on-probability (that is, the probability of producing an output of 1 rather than 0 in the next time step), is given by the logistic activation function

\begin{equation}
p_1(u) = \frac{1}{1 + e^{-u/t_0}}.
\end{equation}
Here, $t_0$ is an effective temperature parameter that has been set to $t_0 = 5$ after numerical optimization. 

\NI For neuron $i$ in the core network, the total synaptic input $u_i(t)$ at time step $t$ consists of three components from other core neurons, from interneurons and from peripheral neurons: 
\begin{eqnarray}
u_i(t) &=& \sum_k W_{ik}^{CC}z_k(t) \,+\, 
           \sum_m W_{im}^{CI}y_m(t) \,+\,
           \sum_n W_{in}^{CP}x_n(t) \,= \nonumber\\
       &=& u_{i,cor}(t) \,+\, 
           u_{i,int}(t) \,+\,
           u_{i,per}(t) \,= \nonumber\\
       &=& u_{i,cor}(t) \,+\, u_{i,emb}(t),       
\end{eqnarray}
where 
\begin{equation}
u_{i,emb}(t)=u_{i,int}(t)+u_{i,per}(t)
\label{EqUEmb}
\end{equation}
denotes the total contribution from the embedding networks.

% *******************************************************

\subsection{Information Flux in the Core Network}
\label{MethInfoFlux}

\NI At any given time step $t$, the global state of the core network is defined by a vector $\mathbf{z}(t)=\left(z_0(t),z_1(t),\ldots\right)$, where $z_n(t)\in\left\{0,1\right\}$ denotes the binary state of neuron $n$. We define the {\bf information flux} within this network as the mutual information $I\left(\mathbf{z}(t);\mathbf{z}(t\!+\!1)\right)$ between two successive global states.

\NI For the numerical evaluation of the mutual information
\begin{equation}
I(X;Y) = \sum_x \sum_y P(x,y) \log_2 \left( \frac{P(x,y)}{P(x) P(y)} \right),
\end{equation}
we simulate long time series of network states and estimate the joint probability distribution $P(x,y)$ by counting pairs of successive states $\left(x=\mathbf{z}(t),\,y=\mathbf{z}(t\!+\!1)\right)$, excluding all terms for which one or more of the involved probabilities are zero.

\NI In a network of $N$ neurons, $P(x,y)$ is represented by a matrix of size $2^N\times2^N$, where each matrix element corresponds to a state pairing that should occur at least $K\approx 10$ times in the time series for sufficient numerical accuracy. Assuming, for simplicity, equally frequent state pairings, the required simulation time scales as $T_{\mathrm{sim}} = K \cdot 2^{2N}$. In our core network with $N = N_c = 10$ and $K = 10$, this yields $T_{\mathrm{sim}} \approx 10^7$ time steps per run. Since we need to perform simulations over many parameter settings, the computation of the full global information flux is not feasible.

\NI Instead, we employ an approximation based on smaller subgroups of neurons. To this end, we select three distinct triplets within the core network: $A=\left\{0,1,2\right\}$, $B=\left\{3,4,5\right\}$, and $C=\left\{6,7,8\right\}$. We then define the {\bf intra-triplet flux $I_{intra}$}   by computing $I_{AA}=I\left(\mathbf{z}_A(t);\mathbf{z}_A(t\!+\!1)\right)$, $I_{BB}$, and $I_{CC}$ independently for each triplet and averaging over the three values. In addition, we define the {\bf inter-triplet flux $I_{inter}$} by computing $I_{AB}=I\left(\mathbf{z}_A(t);\mathbf{z}_B(t\!+\!1)\right)$, as well as $I_{AC}$, $I_{BC}$, $I_{BA}$, $I_{CA}$, and $I_{CB}$, and averaging over these six pairwise values. Finally, we compute the sum of the intra- and inter-triplet flux and use it as a simple general {\bf flux indicator} $f=I_{intra}+I_{inter}$ in the core. This quantity $f$ is presented as a simple number, suppressing the unit of $10^{-3}$ bit. 

% *******************************************************

\subsection{Mutual information of successive activations in a self-coupled noisy Boltzmann neuron}
\label{MethAnalytRR}

\NI We consider a single binary Boltzmann neuron with states
\begin{equation}
x_t \in \{0,1\},
\end{equation}
recursive self-coupling strength $w$, bias $b$, and additive noise $\xi_t$ drawn independently at each time step from a symmetric uniform distribution on $[-a,a]$. The total input at time $t+1$ is therefore
\begin{equation}
u_t = w x_t + b + \xi_t.
\end{equation}
The underlying activation function, which describes the probability of producing an on-state in the next time step, is logistic,
\begin{equation}
p_1(u) = \frac{1}{1+\exp(-u)}.
\end{equation}

\NI Averaging over the uniform noise yields the effective activation probability
\begin{equation}
F(u;a) = \frac{1}{2a}\int_{-a}^{a}p_1(u+\xi)\,d\xi
= \frac{1}{2a}\left[\ln\!\left(1+\exp(u+a)\right)-\ln\!\left(1+\exp(u-a)\right)\right].
\end{equation}
For the two possible previous states, we define
\begin{equation}
A \equiv F(b;a),
\qquad
C \equiv F(w+b;a).
\end{equation}
Hence the transition probabilities are
\begin{align}
P(1|0) &= A, &
P(0|0) &= 1-A, \\
P(1|1) &= C, &
P(0|1) &= 1-C.
\end{align}
The corresponding Markov transition matrix is
\begin{equation}
T =
\begin{pmatrix}
1-A & A \\
1-C & C
\end{pmatrix}.
\end{equation}

\NI Let $\pi_0$ and $\pi_1$ denote the stationary probabilities of states $0$ and $1$. From
\begin{equation}
\pi_1 = \pi_0 A + \pi_1 C,
\qquad
\pi_0 + \pi_1 = 1,
\end{equation}
one obtains
\begin{equation}
\pi_1 = \frac{A}{1-C+A},
\qquad
\pi_0 = \frac{1-C}{1-C+A}.
\end{equation}

\NI The joint probabilities of two successive states,
\begin{equation}
p_{ij} \equiv P(x_t=i,x_{t+1}=j),
\end{equation}
follow as
\begin{align}
p_{00} &= \pi_0(1-A) = \frac{(1-C)(1-A)}{1-C+A}, \\
p_{01} &= \pi_0 A = \frac{A(1-C)}{1-C+A}, \\
p_{10} &= \pi_1(1-C) = \frac{A(1-C)}{1-C+A}, \\
p_{11} &= \pi_1 C = \frac{AC}{1-C+A}.
\end{align}

\NI The mutual information between two successive activations is
\begin{equation}
I(x_t;x_{t+1}) =
\sum_{i,j\in\{0,1\}} p_{ij}
\log_2\!\left(\frac{p_{ij}}{\pi_i\pi_j}\right).
\end{equation}
This can be rewritten in terms of entropy and conditional entropy as
\begin{equation}
I(x_t;x_{t+1}) = H(x_{t+1}) - H(x_{t+1}|x_t),
\end{equation}
where
\begin{equation}
H(x_{t+1}) = -\sum_{j\in\{0,1\}} \pi_j \log_2 \pi_j,
\end{equation}
and
\begin{equation}
H(x_{t+1}|x_t) =
-\sum_{i,j\in\{0,1\}} p_{ij}\log_2 P(x_{t+1}=j|x_t=i).
\end{equation}

\NI Introducing the binary entropy function
\begin{equation}
h_2(p) = -p\log_2 p -(1-p)\log_2(1-p),
\end{equation}
these quantities become
\begin{equation}
H(x_{t+1}) = h_2(\pi_1),
\end{equation}
and
\begin{equation}
H(x_{t+1}|x_t) = \pi_0 h_2(A) + \pi_1 h_2(C).
\end{equation}
Hence,
\begin{equation}
I(x_t;x_{t+1}) = h_2(\pi_1) - \pi_0 h_2(A) - \pi_1 h_2(C).
\end{equation}

\NI Substituting the stationary probabilities finally yields
\begin{equation}
I(x_t;x_{t+1})
=
h_2\!\left(\frac{A}{1-C+A}\right)
-
\frac{(1-C)h_2(A)+A\,h_2(C)}{1-C+A},
\end{equation}
with
\begin{equation}
A = F(b;a),
\qquad
C = F(w+b;a),
\end{equation}
and
\begin{equation}
F(u;a)=
\frac{1}{2a}\left[\ln\!\left(1+\exp(u+a)\right)-\ln\!\left(1+\exp(u-a)\right)\right].
\end{equation}

\NI Thus, the mutual information is obtained explicitly as a function of the self-coupling strength $w$, the bias $b$, and the half-width $a$ of the added uniform noise.

% *******************************************************

\subsection{Neurons with adaptive bias}
\label{MethAdaptBias}

\NI The output entropy of a binary Boltzmann neuron is maximal when the on-probability is $p_1\is1/2$. In this case, the time-averaged output of the neuron satisfies $\left\langle z_i(t)\right\rangle_t = 1/2$. To drive the neuron toward this maximum-entropy operating point, we let the bias $b_i(t)$ adapt dynamically according to

\begin{equation}
b_i(t) = b_i(t\!-\!1) - \epsilon \left(z_i(t)\!-\!\frac{1}{2}\right),
\end{equation}

\NI where $\epsilon\is0.01$ is a small adaptation rate constant. This update rule implements a local negative feedback mechanism: if the neuron is active more often than desired ($z_i(t)>1/2$), the bias is decreased, whereas for too infrequent activation ($z_i(t)<1/2$), the bias is increased. Over time, this drives the average activity toward $1/2$.

% *******************************************************
\section{Results}
% *******************************************************

\subsection{Optimizing the Network's Operating Point}

\begin{figure}[ht!]
\centering
\includegraphics[width=0.68\linewidth]{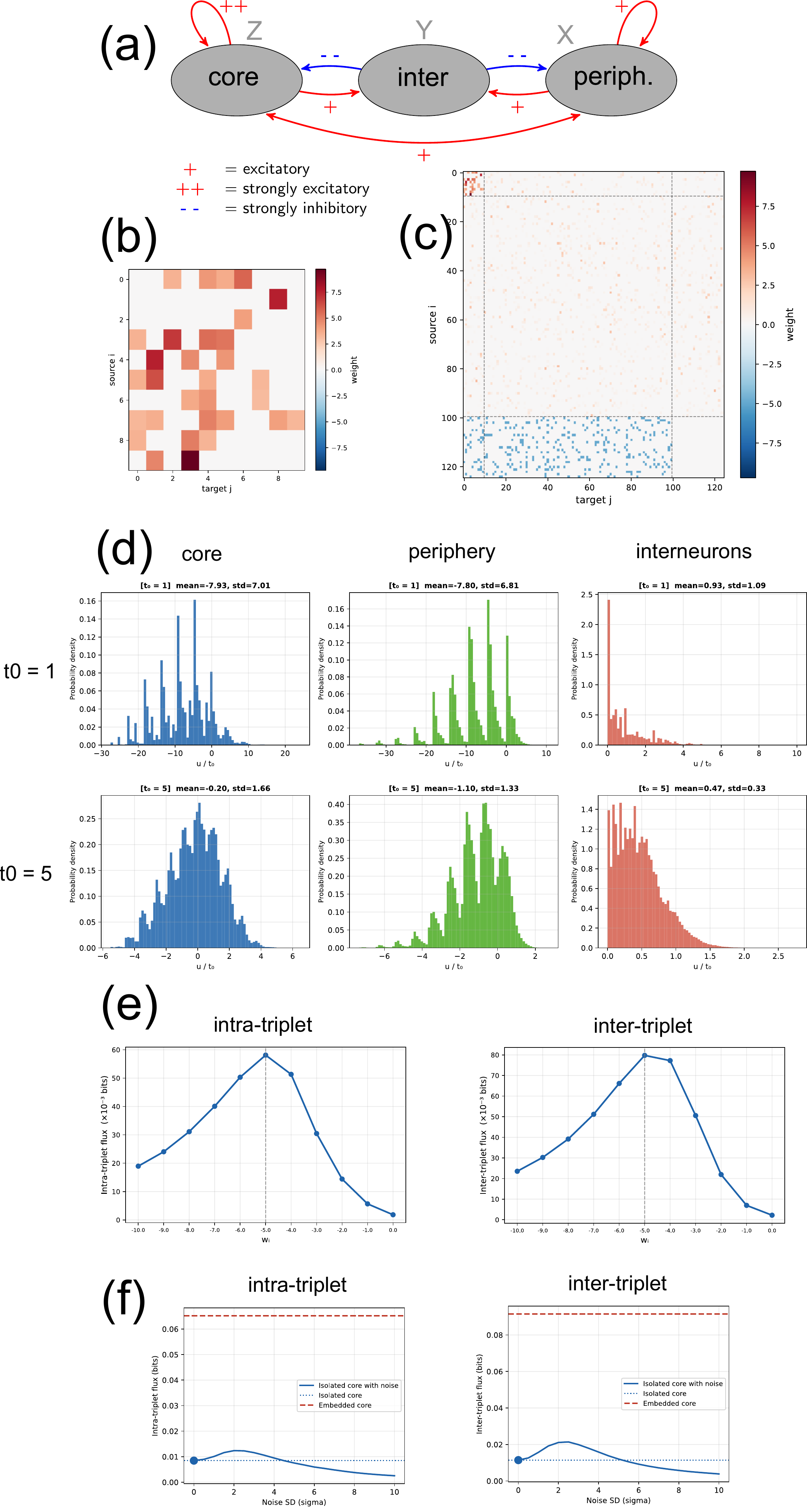}
\caption{
{\bf The model system.}
\textbullet$\;${\bf(a)} Three neural sub-populations and their connections.
\textbullet$\;${\bf(b)} Connection weight matrix of isolated core.
\textbullet$\;${\bf(c)} Connection weight matrix of the full network.
\textbullet$\;${\bf(d)} Distributions of scaled input sums $p(u/t_0)$ in the three sub-populations (using $w_i\is-5$), for two different neural temperatures $t_0\is1$ and $t_0\is5$. 
\textbullet$\;${\bf(e)} Intra- and inter-triplet information flux in the embedded core as a function of the inhibitory weight $w_i$.
\textbullet$\;${\bf(f)} Intra- and inter-triplet information flux in the isolated core (using $t_0\is5$) as a function of externally injected noise strength $\sigma$. Red dashed line corresponds to embedded core with $w_i\is-5$, blue dotted line to isolated core without noise.
} 
\label{Fig_A}
\end{figure}

\NI Our full Embedded Core Model (ECM), described in detail in the Method section \ref{MethECM} and illustrated schematically in Fig.\ref{Fig_A}(a), consists of three interconnected populations: a core network of $N_c\is10$ neurons, a periphery network of $N_p\is90$ neurons, and a group of $N_i\is25$ interneurons (See Fig.\ref{Fig_A}(b) for the connection weight matrix of the core and (c) for that of the whole system). The peripheral and interneuron populations are considered the 'embedding' networks of the core.

\NI The overall connectivity pattern and the log-normal distribution of excitatory weights are intended to reflect measured properties of cortical microcircuits \cite{song2005highly}. However, not all parameters required for our simulations are constrained by experiment. In particular, the relative strength of inhibitory weights is not specified in the biological data and must therefore be treated as a free parameter. In addition, because we use highly simplified model neurons, a global rescaling of all synaptic weights is required. The present section describes these preparatory steps, which are necessary to place the network into a suitable operating regime.

\NI Throughout this paper, neurons are modeled as Boltzmann units, for which the probability of firing in the next time step is given by a logistic function of the total input sum $u$. At $u\is0$, the firing probability is $p_1(u\is0)\is1/2$, and in the interval $u\in\left[-1,+1\right]$ around this point the logistic function is still approximately linear. By contrast, for larger magnitudes, roughly $|u|>5$, the response becomes strongly saturated. For effective network dynamics, the neurons should operate neither in an almost purely linear regime nor in a permanently saturated one. To control this, we introduce a neural temperature parameter $t_0$, such that the argument of the logistic function is given by $u/t_0$.

\NI To identify a suitable operating regime, we first consider the unscaled network, corresponding to $t_0\is1$. In this case, the values of $u/t_0$ in the core, periphery, and interneuron populations are distributed far too broadly and extend deep into the saturation regime (Fig.\ref{Fig_A}(d), upper row).

\NI We therefore increased the temperature parameter systematically and found that for $t_0\is5$ all three populations operate in a much more suitable range (Fig.\ref{Fig_A}(d), lower row). This value will therefore be used throughout the remainder of the paper.

\NI A second open parameter is the relative inhibitory connection strength $w_i$, which is not provided in \cite{song2005highly}. Since our aim is to maximize the information flux within the core network, we choose $w_i$ such that the mutual information between successive core states becomes maximal.

\NI As described in \ref{MethInfoFlux}, the mutual information is evaluated, for practical reasons, only for groups of three neurons (``triplets''), so that the maximum possible information flux is $I_{opt}\is3$ bit. In the following, we distinguish between intra-triplet and inter-triplet information flux. For simplicity, we also add the two measures and use their sum as a general flux indicator $f$. Although originally of order $10^{-3}\,\mathrm{bit}$, we present this quantity $f$ in the following as a dimensionless number. 

\NI Sweeping the inhibitory strength $w_i$ across a range of values reveals a clear optimum at $w_i\is-5$. At this point, the information flux in the core reaches about $0.057$ bit for the intra-triplet contribution and about $0.082$ bit for the inter-triplet contribution, corresponding to a flux indicator of $f\is139.3$. Although these values remain far below the theoretical maximum of $I_{thm}\is3$ bit, the relevant question is how they compare with the corresponding values of isolated core networks that lack the support of the periphery and interneuron populations.

% *******************************************************

\subsection{Isolated Core Network}

\NI In the next step, we therefore analyze the core network in complete isolation from the other two populations. Remarkably, the information flux in the core then drops to $0.009$ for the intra-triplet case and to $0.014$ for the inter-triplet case, corresponding to a flux indicator of $f\is23$. This clearly demonstrates that the surrounding network exerts a strong influence on the core dynamics, leading to a substantial enhancement of information flux.

\NI The extremely small information fluxes in the isolated core are a direct consequence of the strong and purely excitatory internal coupling within this subnetwork. In the absence of regulating signals from the outside, runaway positive feedback drives the core into a dynamical regime in which most neurons remain in the on-state, $z_i(t)\is1$, for most time steps. Indeed, when averaged over all neurons and time steps, the global on-probability in the isolated core is about $p_{on}\is0.89$.

\NI The mutual information between successive time steps is maximal when the system visits all of its possible states with comparable probability, corresponding to high entropy, while the transitions between these states remain largely deterministic. In the isolated core, however, the first condition is strongly violated, since the dynamics spend most of the time in a small subset of states in which nearly all neurons are active. This confinement to a restricted region of phase space drastically reduces the entropy, which in turn limits the attainable information flux.

\NI It is clear that the supporting peripheral and interneuron populations can mitigate the trapping of the core to a significant extent. In principle, this effect could arise from a sophisticated regulatory feedback between the core and the surrounding network. Alternatively, it could result from complex forms of effective noise, such as signals with non-Gaussian distributions or temporal correlations. The actual underlying mechanism will later be revealed by a systematic reverse-engineering analysis of the ECM. Before that, we investigate white Gaussian noise as an alternative means of raising the information flux in the core.

% *******************************************************

\subsection{Noise-enhanced Information Flux}

\NI As established in earlier work \cite{krauss2019recurrence,metzner2024recurrence}, the information flux in RNNs that are confined to small regions of state space can be increased by injecting an optimal amount of noise. To this end, we add independent white Gaussian noise to each of the $N_t$ neurons in the isolated core. As the standard deviation $\sigma$ of the noise is increased, the information flux initially rises, reaches around $\sigma\is2$ a maximum of $0.012$ for the intra-triplet case and $0.022$ for the inter-triplet case, corresponding to a flux indicator of $f\is34.0$. For larger noise strength, the information flux decreases again to values even below those of the noise-free system. This characteristic behavior is referred to as Recurrence Resonance. 

\NI Taken together, these results show that adding white Gaussian noise to the isolated core substantially enhances the information flux from $f\is23.0$ to $f\is34.0$. However, this enhancement remains clearly below the level achieved when the core is embedded within the full network comprising periphery and inhibitory populations, where the flux indicator is at $f\is139.3$. For direct comparison, Table~\ref{tab:flux_comparison} summarizes the measured values for all configurations.

\begin{table}[h]
\centering
\renewcommand{\arraystretch}{1.2}
\begin{tabular}{|l|c|c|c|}
\hline
\textbf{Network configuration} & \textbf{Intra-triplet flux (bit)} & \textbf{Inter-triplet flux (bit)} & \textbf{Flux indicator} \\
\hline
Isolated core 
& $0.009$ 
& $0.014$
& $23.0$ \\
\hline
Noise-assisted core 
& $0.012$ 
& $0.022$
& $34.0$ \\
\hline
Embedded core 
& $0.057$ 
& $0.082$
& $139.3$ \\
\hline
Indiv.\ biases ($\bar{u}_{i,\mathrm{emb}}$)
& $0.042$
& $0.067$
& $109.5$ \\
\hline
Indiv.\ biases $+$ normal noise
& $0.011$
& $0.019$
& $30.3$ \\
\hline
Uniform optimized biases
& $0.075$
& $0.112$
& $187.4$ \\
\hline
Indiv.\ optimized biases
& $0.083$
& $0.121$
& $203.9$ \\
\hline
\end{tabular}
\caption{Comparison of information flux across different network configurations. The noise-assisted case corresponds to the peak of recurrence resonance. The flux indicator is the sum of intra- and inter-triplet flux, expressed as a dimensionless number (original units: $\times 10^{-3}$\,bit).}
\label{tab:flux_comparison}
\end{table}

% *******************************************************

\subsection{Analytical Theory of a Noisy Boltzmann Neuron}

\begin{figure}[ht!]
\centering
\includegraphics[width=0.9\linewidth]{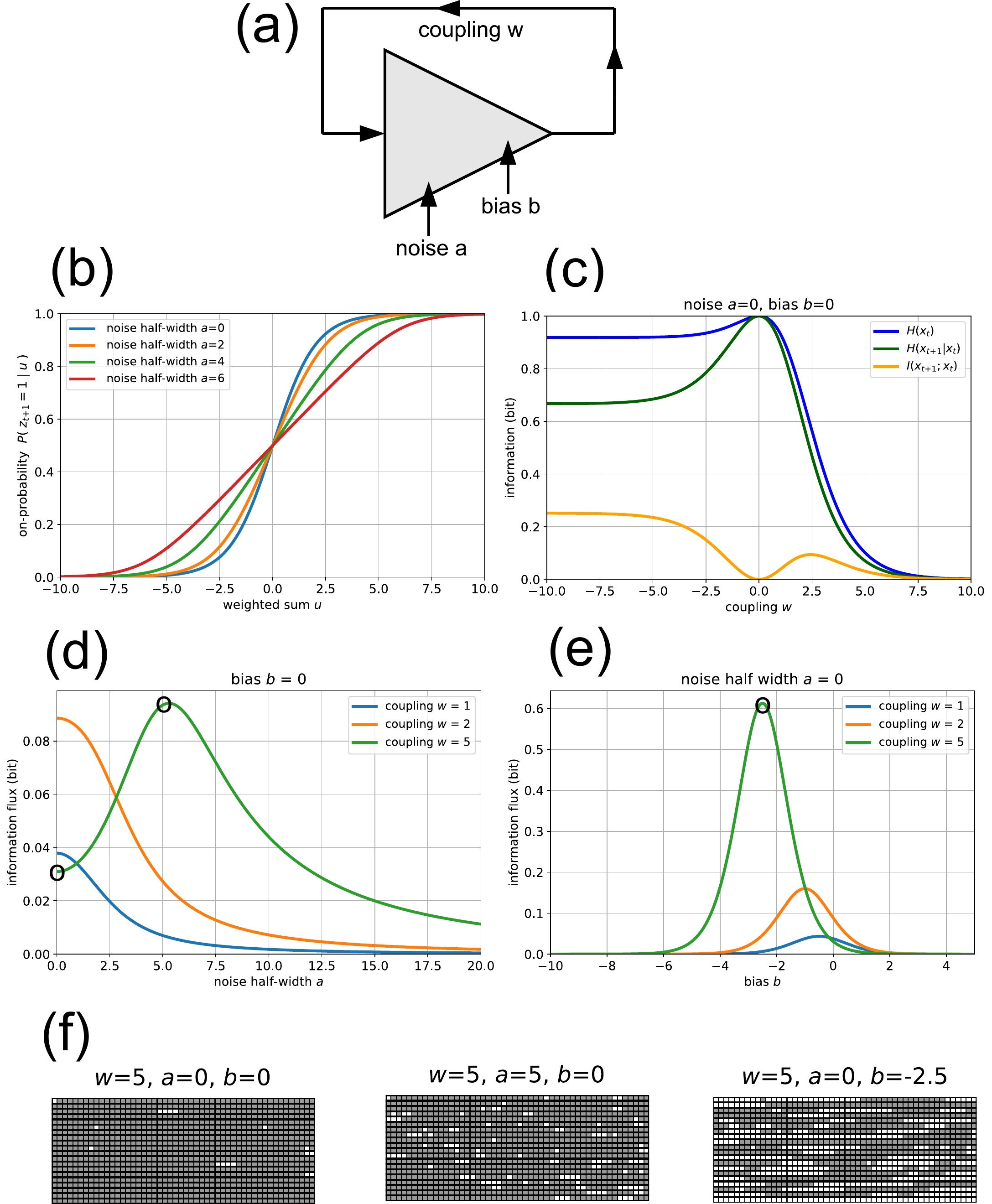}
\caption{
{\bf Analytical theory of a single noisy Boltzmann neuron.}
\textbullet$\;${\bf(a)} Circuit diagram.
\textbullet$\;${\bf(b)} Effective response function for different noise strengths $a$.
\textbullet$\;${\bf(c)} Entropy $H(x_t)$, conditional entropy $H(x_{t\!+\!1}|x_t)$ and mutual information $I(x_{t\!+\!1};x_t)$ (information flux) as a function of coupling strength $w$, in a system without noise and without bias.
\textbullet$\;${\bf(d)} Information flux as a function of noise strength $a$ for three different recurrent coupling strengths $w$, without applied bias.
\textbullet$\;${\bf(e)} Information flux as a function of bias $b$ for three different recurrent coupling strengths $w$, without applied noise.
\textbullet$\;${\bf(f)} Simulated time series of 1000 successive binary activations, produced by a strongly coupled ($w\is5$) neuron, in different parameter regimes. Left: No noise and no bias. Middle: Optimal noise. Right: Optimal bias. The corresponding operating points are marked by black circles in (c) and (d).
} 
\label{Fig_B}
\end{figure}

\NI At this point, it is useful to note that Recurrence Resonance can also be analyzed analytically in simplified systems. To illustrate this, we consider a minimal model consisting of a single Boltzmann neuron with self-coupling $w$ and an optional bias $b$ (Fig.\ref{Fig_B}(a)). For analytical convenience, Gaussian noise is replaced by noise drawn from a uniform distribution over the interval $[-a,+a]$, which effectively broadens the neuronal response function $p_1(u)$ (Fig.\ref{Fig_B}(b)). Under these assumptions, the information flux $I$ can be calculated explicitly as a function of the coupling strength $w$, the bias $b$, and the noise amplitude $a$ (see Methods section \ref{MethAnalytRR}).

\NI It is instructive to first consider the dynamics of the neuron without noise and without bias as a function of the self-coupling strength $w$ (Fig.\ref{Fig_B}(c)). The information flux, measured by the mutual information $I(x_{t\!+\!1};x_t)\is H(x_t)-H(x_{t\!+\!1}|x_t)$, is the difference between the entropy $H(x_t)$, which quantifies the overall variability of the states reached by the system, and the conditional entropy $H(x_{t\!+\!1}|x_t)$, which quantifies the remaining uncertainty in the next state once the current state is known.

\NI For strongly negative coupling $w$, the neuron tends to oscillate. However, a perfectly regular alternation between 1- and 0-states cannot be achieved without bias, because after each 0-state there is still only a probability of $p\is1/2$ that the next state will be 1. This built-in indeterminism after 0-states is reflected in a relatively large value of the conditional entropy (green line in Fig.\ref{Fig_B}(c)). Although the entropy (blue line) is close to its optimal value of one, the resulting information flux (orange line) remains limited to about $I\approx0.25$.

\NI At $w\is0$, the Boltzmann neuron behaves like an unbiased binary random-number generator, with statistically independent states in successive time steps. Due to this completely indeterministic behavior, the information flux drops to zero.

\NI As $w$ becomes more positive, longer sequences of 1-states start to occur. This increasingly deterministic behavior causes the conditional entropy to drop from one toward zero. However, after each 0-state the neuron still switches randomly between the two possible successor states, so that long 0-sequences remain very unlikely compared to long 1-sequences. Consequently, the entropy also drops toward zero, although at a slightly different rate than the conditional entropy. As a result, the information flux shows a characteristic peak around $w\approx2.5$. For very large positive $w$, the information flux again approaches zero, because it is bounded from above by the entropy of the state distribution.

\NI The analytical treatment also shows that a resonance-like peak in the information flux as a function of the noise amplitude $a$ occurs only in strongly coupled systems, where runaway dynamics would otherwise confine the neuron to a small subset of states (Fig.\ref{Fig_B}(d), green curve). By contrast, for weaker coupling strengths $w$, the information flux decreases monotonically with increasing noise amplitude $a$ (Fig.\ref{Fig_B}(d), blue and orange curves).

\NI The same analytical theory further shows that, instead of adding noise, applying a constant bias can be a highly effective way to increase the information flux (Fig.\ref{Fig_B}(e)), even in weakly coupled systems.

\NI Simulated activation time series of a Boltzmann neuron with strongly excitatory self-coupling (Fig.\ref{Fig_B}(f)) support this interpretation. Without any countermeasure, that is, in the absence of both noise and bias, the neuron generates very long sequences of on-states, only occasionally interrupted by short off-periods. This strong imbalance corresponds to a very small information flux of only $I\approx0.03$ (Fig.\ref{Fig_B}(f), left panel).

\NI Adding an optimal amount of noise increases both the frequency and the duration of the off-phases and thereby raises the information flux to $I\approx0.095$ (Fig.\ref{Fig_B}(f), middle panel). By contrast, applying an optimal bias produces a much more balanced alternation between on- and off-states and leads to a dramatic increase of the information flux to $I\approx0.6$ (Fig.\ref{Fig_B}(f), right panel).

\NI In this system, the theoretical maximum of the information flux is $I_{\mathrm{thm}}\is1$. In principle, this limit could be approached in systems with even stronger self-coupling combined with an optimal bias.

% *******************************************************

\subsection{Simulated Lesion Experiments}

\begin{figure}[ht!]
\centering
\includegraphics[width=1\linewidth]{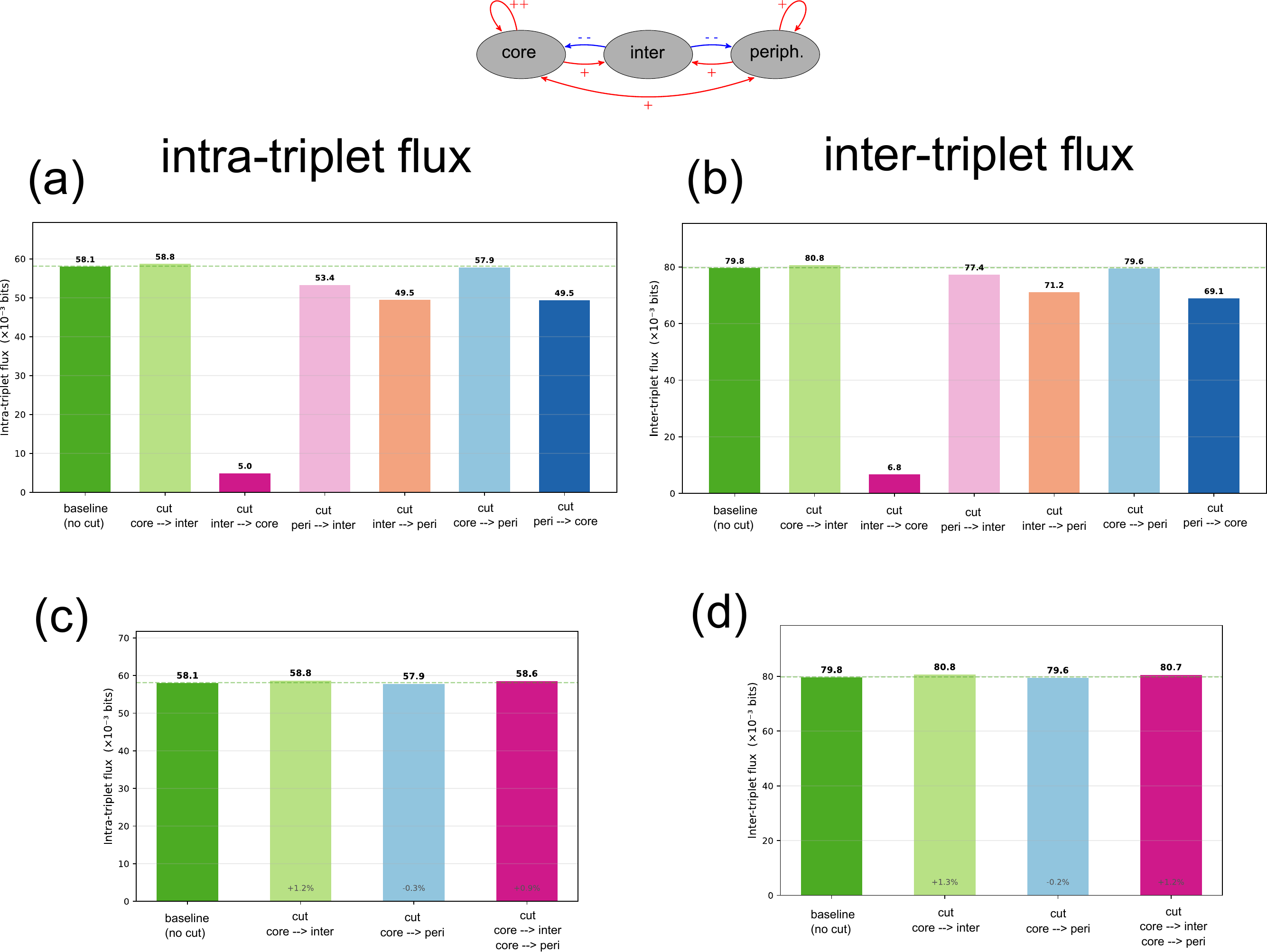}
\caption{
{\bf Simulated lesion experiments.} Specific directional connections between the three neural populations (compare circuit diagram on top) are cut and the effect on the information flux in the core network is investigated. The left two columns (a,c) show the effect on the intra-triplet flux and right two columns (b,d) on the inter-triplet flux. For the final magenta bars in (c) and (d), all outgoing connections from the core to the two other populations have been cut simultaneously. 
} 
\label{Fig_C}
\end{figure}

\NI To uncover the mechanism by which the embedding network enhances information flux in the core, we next perform simulated lesion experiments. In these experiments, specific directional connections between the three neural populations are completely removed, that is, set to zero (Fig.\ref{Fig_C}). We then quantify how these lesions affect both the intra-triplet and inter-triplet flux measured within the core network.

\NI Surprisingly, removing all outgoing connections from the core to the peripheral population (Fig.\ref{Fig_C}(a,b), dark blue) or from the core to the interneurons (Fig.\ref{Fig_C}(a,b), light green) has almost no effect on the information flux in the core. This conclusion is further supported by simultaneously cutting both sets of outgoing projections (Fig.\ref{Fig_C}(c,d), magenta), which again leaves the flux nearly unchanged. The flux-enhancing effect of the embedding network therefore cannot be explained by a closed regulatory feedback loop between the core and its surrounding populations.

\NI In contrast, removing the connections from the interneurons to the core causes a drastic breakdown of information flux (Fig.\ref{Fig_C}(a,b), magenta). This shows that inhibitory input from the interneurons is the dominant factor driving the core out of the restricted region of state space in which it would otherwise remain trapped. Removing the connections from the peripheral population to the core also reduces the information flux, although much less strongly (Fig.\ref{Fig_C}(a,b), dark blue). This result is noteworthy because the periphery provides only excitatory input to the core.

\NI Removing the connections between the peripheral population and the interneurons, in either direction, also leads to a small but significant reduction of information flux in the core (Fig.\ref{Fig_C}(a,b), pink and orange). This suggests that the mutual interactions between these two surrounding populations help maintain a dynamical regime from which both can deliver beneficial signals to the core.

\NI In summary, the lesion experiments show that the enhancement of information flux is not based on regulatory feedback from the core to the embedding network. Instead, the peripheral population and the interneurons provide input signals to the core whose statistical properties are largely independent of the momentary state of the core itself. Whether these signals should be interpreted primarily as noise-like perturbations or as effective biases will be clarified in the following experiments.

% *******************************************************

\subsection{Neural Activations and Pairwise Mutual Information}

\begin{figure}[ht!]
\centering
\includegraphics[width=0.88\linewidth]{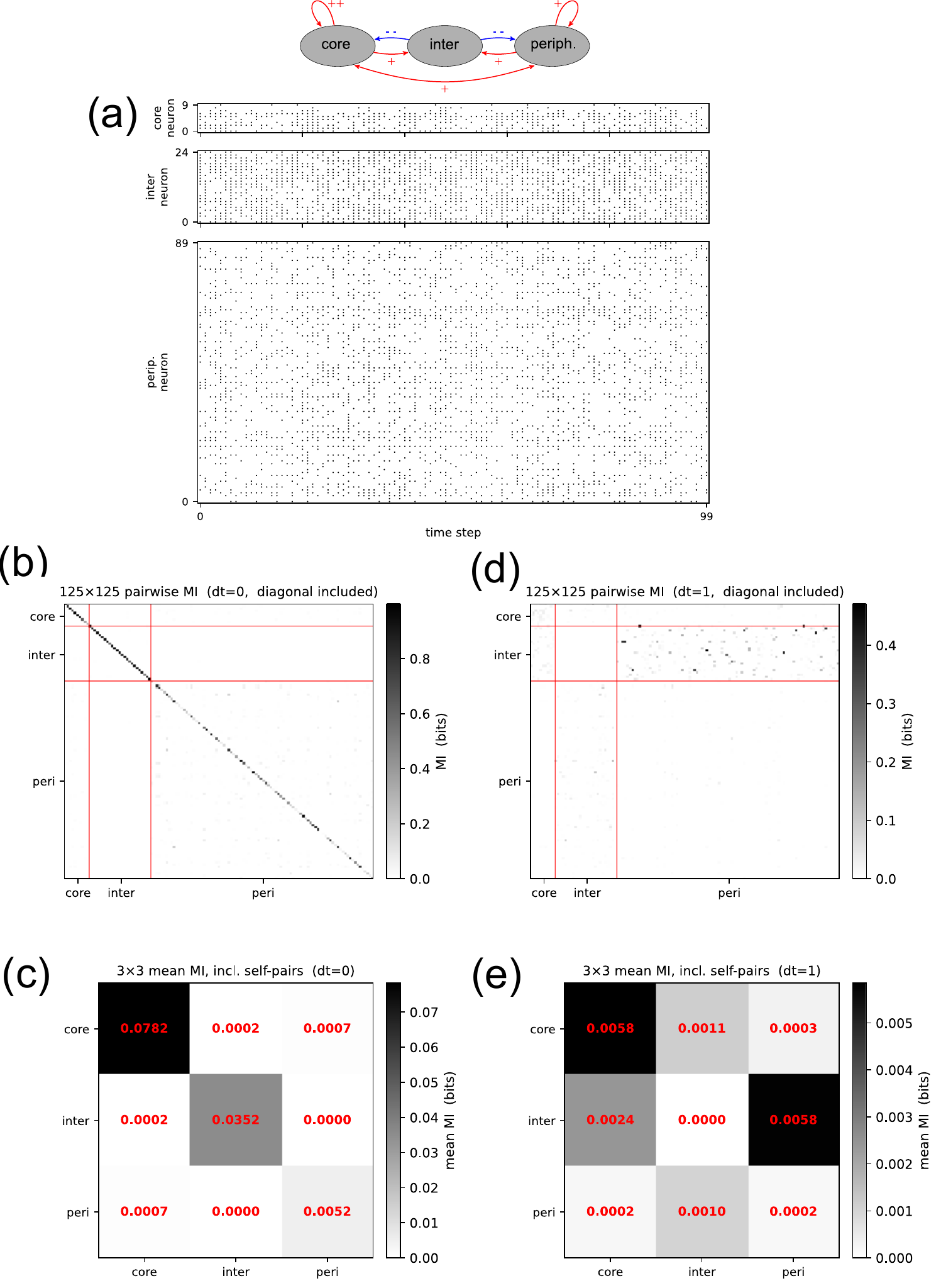}
\caption{
{\bf Activation statistics in the full embedded core network.}
\textbullet$\;${\bf(a)} Activations of all 125 Boltzmann neurons over 100 successive time steps, with active states marked by black dots.
\textbullet$\;${\bf(b)} Matrix of instantaneous pairwise mutual information for the full $125\times125$ neuron system. 
\textbullet$\;${\bf(c)} Reduced mutual information matrix obtained from panel (b) by averaging over all individual neuron pairs within each combination of the three sub-populations.
\textbullet$\;${\bf(d)} Matrix of time-delayed pairwise mutual information ($dt=1$) for the full $125\times125$ neuron system.
\textbullet$\;${\bf(e)} Reduced time-delayed mutual information matrix obtained from panel (d) by averaging over all individual neuron pairs within each combination of the three sub-populations. For example, the entry in row ``inter'' and column ``peri'' represents the average information flux from interneurons to peripheral neurons.
}
\label{Fig_D}
\end{figure}

\NI For a direct visualization of the network dynamics, we plot the binary activation time series of all 125 neurons over 100 successive time steps (Fig.\ref{Fig_D}(a)). It is immediately apparent that neural on-probabilities differ across the three sub-populations. Numerical evaluation of the ensemble averages yields $p_1\is0.288$ in the core, $p_1\is0.125$ in the peripheral population, and $p_1\is0.683$ in the interneuron population.

\NI Two target neurons $x$ and $y$ that both receive input from the same source neuron $z$ can exhibit instantaneous statistical dependencies in their activation time series. These dependencies can be quantified by the mutual information $I(x_t;y_t)$, which ranges between 0 and 1 bit for binary neurons. In particular, within the core, which possesses strong recurrent internal connections, we therefore expect to find pairs of neurons with non-zero simultaneous mutual information. To test this hypothesis, we first compute the full $125\times125$ matrix of instantaneous pairwise mutual information in the system (Fig.\ref{Fig_D}(b)). The diagonal elements of this matrix dominate, but as they correspond to the entropy of individual neurons, they are not all equal to one. In a next step, we reduce the full matrix to a $3\times3$ matrix by averaging over all neuron pairs within each of the nine combinations of sub-populations (Fig.\ref{Fig_D}(c)). Indeed, we find the strongest simultaneous mutual information within the core (0.0782 bit). Remarkably, the second largest value is found within the interneuron population (0.0352 bit), and only the third largest within the peripheral population (0.0052 bit). The relatively large statistical dependencies between instantaneous activations of the interneurons are noteworthy, given the role of the interneuron population in enhancing the information flux within the core.

\NI Equally informative is the time-delayed ($\Delta t\is1$) mutual information $I(A_t;B_{t\!+\!1})$ between sub-populations $A$ and $B$ (or within the same population), as it quantifies the predictive influence of $A_t$ on $B_{t\!+\!1}$. In the following, we use this quantity as an effective proxy for the information flux from $A$ to $B$. Again, we first compute the full $125\times125$ matrix (Fig.\ref{Fig_D}(d)). Overall, the diagonal is not dominant in this matrix, indicating that the next state of most neurons is not primarily determined by their own previous state. Only within the core sub-matrix do we observe some relatively strong diagonal entries. Next, we again derive the reduced $3\times3$ matrix (Fig.\ref{Fig_D}(e)) by averaging over neuron pairs. We find that the strongest information flux occurs recursively within the core (0.0058 bit), and with an equal amount from the interneurons to the peripheral population (0.0058 bit), followed by the flux from the interneurons to the core (0.0024 bit).

% *******************************************************

\subsection{Signals from Embedding Networks to the Core}

\NI Next, we analyze the statistical properties of the signals that core neurons receive from the embedding network, that is, from the interneurons and the peripheral population.

\begin{figure}[ht!]
\centering
\includegraphics[width=1\linewidth]{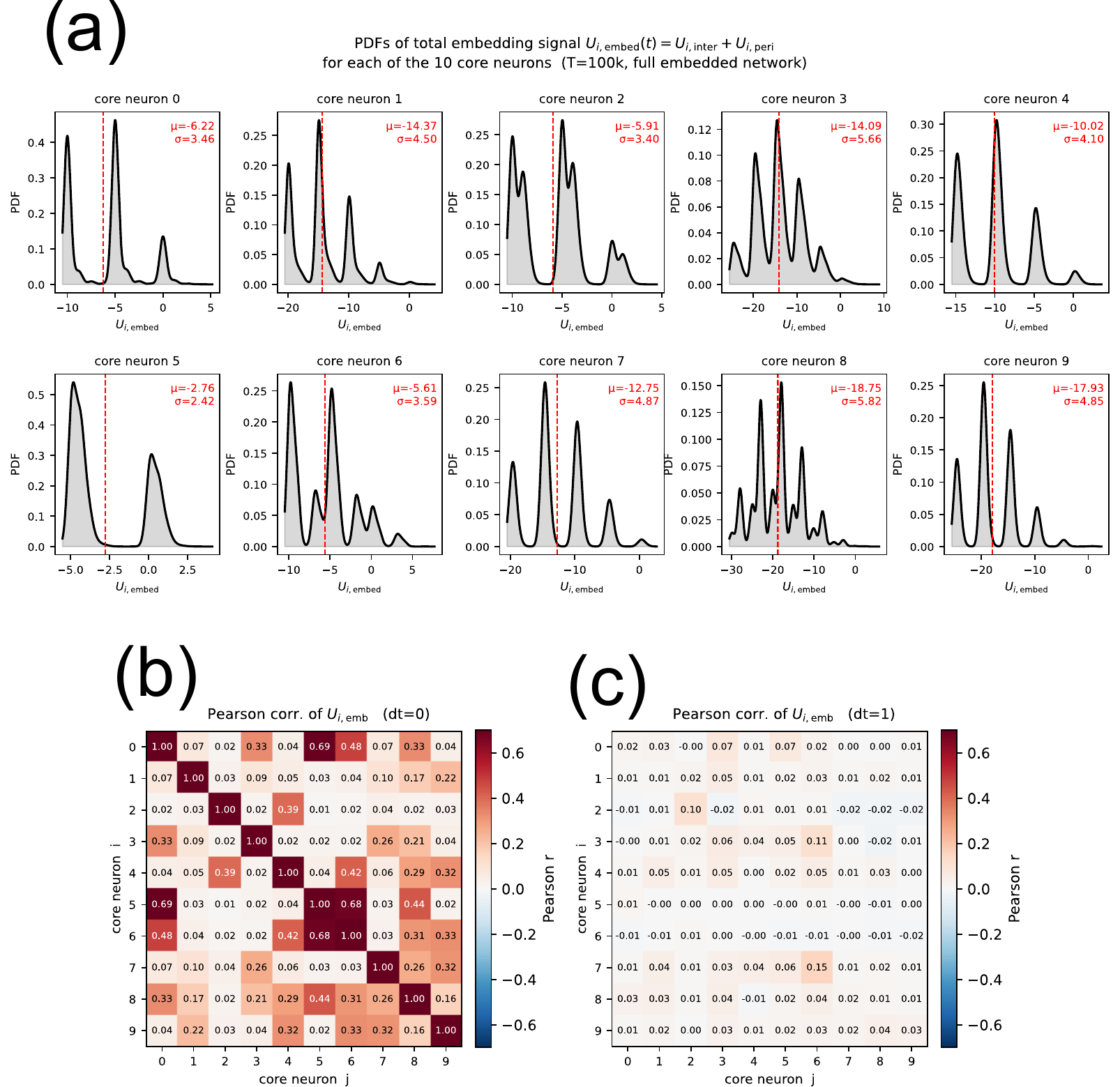}
\caption{
{\bf Analysis of Signals from the Embedding Networks to the Core.}
\textbullet$\;${\bf(a)} Probability density distributions of the total synaptic input $u_{i,\mathrm{emb}}(t)$ from the embedding networks, that is, from the interneurons and the peripheral population, shown separately for each core neuron $i$. Continuous distributions were obtained using Gaussian kernel density estimation with a kernel width of $0.1$ times the standard deviation of the underlying data. Red vertical dashed lines indicate the mean values of $u_{i,\mathrm{emb}}(t)$. The corresponding mean $\mu$ and standard deviation $\sigma$ are additionally displayed within each sub-panel.
\textbullet$\;${\bf(b)} Matrix of pairwise instantaneous Pearson correlations between the embedding inputs $u_{i,\mathrm{emb}}(t)$ and $u_{j,\mathrm{emb}}(t)$ received by different core neurons.
\textbullet$\;${\bf(c)} Like (b), but for time-lagged correlations between $u_{i,\mathrm{emb}}(t)$ and $u_{j,\mathrm{emb}}(t+1)$.
}
\label{Fig_E}
\end{figure}

\NI For this purpose, we record at every time step $t$ the contribution $u_{i,\mathrm{emb}}(t)$ from the embedding network to the total synaptic input of core neuron $i$, as defined in Eq.\ref{EqUEmb}. From the resulting time series, we then compute the probability density functions $p(u_{i,\mathrm{emb}})$ separately for each of the 10 neurons in the core (Fig.\ref{Fig_E}(a)).

\NI Each core neuron receives input from only a small number of neurons in the embedding network. Since the source neurons are binary and their outputs are weighted by fixed synaptic strengths, the resulting input $u_{i,\mathrm{emb}}(t)$ can take only a finite set of discrete values. To obtain a continuous representation of the corresponding probability densities, we apply kernel density estimation with a small kernel width. Nevertheless, the underlying discreteness of the possible input values remains clearly visible in all cases.

\NI The mean $\mu$ and standard deviation $\sigma$ of each distribution $p(u_{i,\mathrm{emb}})$ can be interpreted, to first approximation, as an effective bias and an effective noise amplitude injected into each core neuron by the embedding network. We find that all biases are negative and that, in every case, their magnitude exceeds the corresponding noise amplitude. This raises the question of which of these two effects plays the dominant role in enhancing the information flux within the core network: the bias, which shifts neurons toward a more favorable operating regime, or the noise, which can increase information flux via the recurrence resonance effect.

\NI Before addressing this question, it is important to clarify whether the fluctuations of $u_{i,\mathrm{emb}}(t)$ can be regarded as white noise, and whether the signals from the embedding network act independently on different core neurons. To this end, we compute the pairwise Pearson correlations between the input signals of the core neurons, both at zero lag ($\Delta t\is0$, Fig.\ref{Fig_E}(b)) and with a one-step delay ($\Delta t\is1$, Fig.\ref{Fig_E}(c)).

\NI In the matrix of instantaneous Pearson correlations (b), we observe a considerable number of off-diagonal elements with significantly positive values, indicating that the signals from the embedding networks to different core neurons are not entirely statistically independent. This is expected, since the sets of embedding neurons projecting to different core neurons can partially overlap. By contrast, all pairwise Pearson correlations at time lag $\Delta t\is1$ (c) are very small, indicating that the signals from the embedding networks can be approximated as temporally uncorrelated (white) noise.

% *******************************************************

\subsection{Relevance of Biases and Fluctuations}

\NI Next, we address whether the flux-enhancing effect of the embedding network signals to the core neurons is mainly caused by constant biases or by temporal fluctuations.

\NI For this purpose, we first replace the signals $u_{i,\mathrm{emb}}(t)$ received by each core neuron $i$ from the embedding network by their temporal averages $\overline{u}_{i,\mathrm{emb}}$ (red dashed lines in Fig.\ref{Fig_E}(a)). Thus, the core neurons receive only constant, neuron-specific biases instead of temporally varying control signals.

\NI As a reference, recall that the flux indicator is $f\is139.3$ when the core receives the full signals from the embedding network. After replacing the full signals by constant biases, the flux indicator drops to $f\is109.5$. This shows that constant bias signals alone can explain a large part of the positive effect of the embedding network on the information flux in the core. At the same time, the remaining difference indicates that temporal fluctuations of $u_{i,\mathrm{emb}}(t)$, and thus Recurrence Resonance effects, also contribute.

\NI As a further experiment, we restore random fluctuations around the biases $\overline{u}_{i,\mathrm{emb}}$, but in the simplified form of normally distributed, independent white noise with the same neuron-specific standard deviations as the original complex signals $u_{i,\mathrm{emb}}(t)$. Interestingly, instead of bringing the flux indicator back close to $f\is139.3$, this intervention reduces it to only $f\is30.3$. This indicates that the non-Gaussian probability density distributions of the original embedding signals, as shown in Fig.\ref{Fig_E}(a), are somehow much more effective than Gaussian noise to promote the information flux in the core.

% *******************************************************

\subsection{Optimization of Biases}

\begin{figure}[ht!]
\centering
\includegraphics[width=1\linewidth]{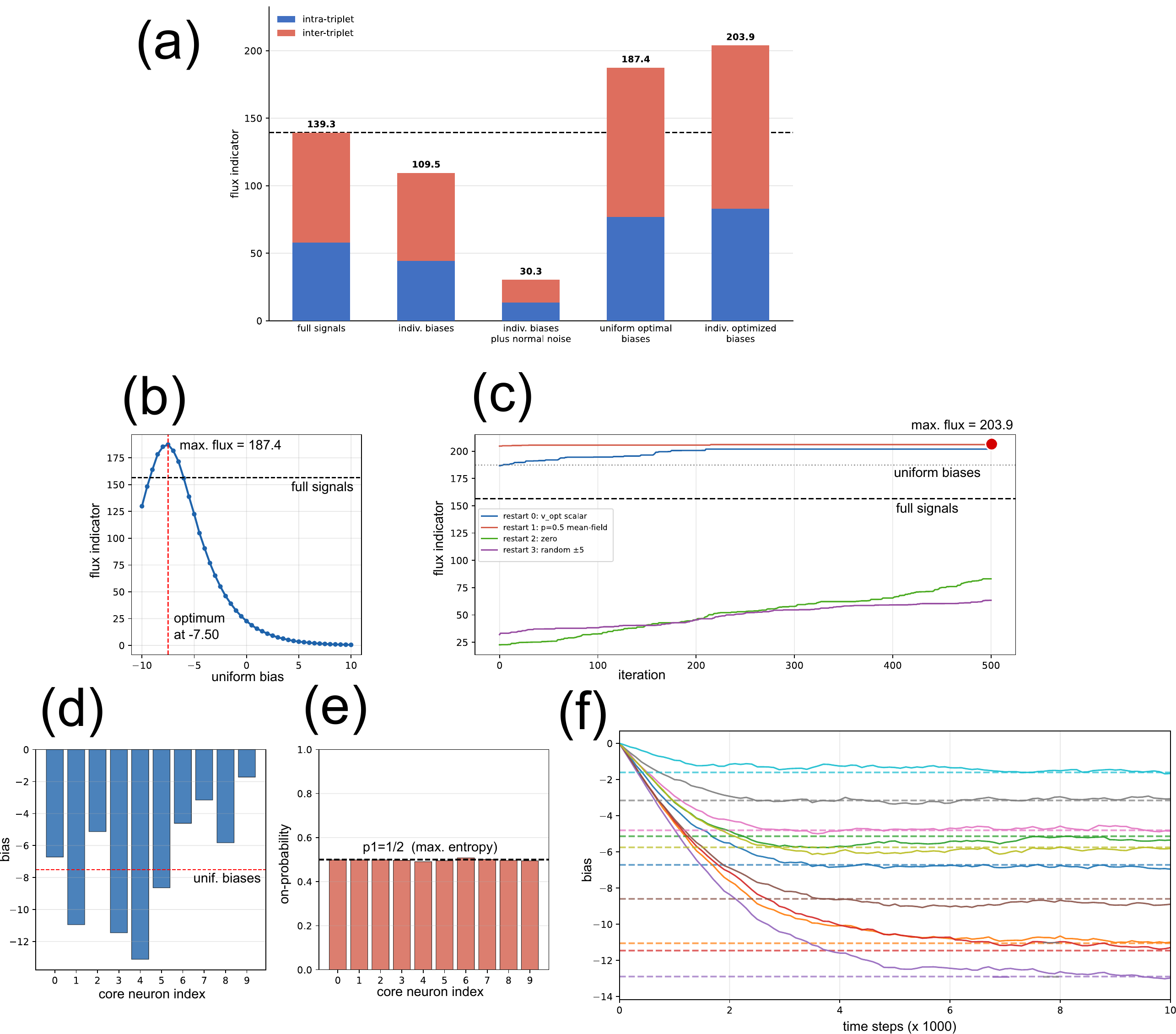}
\caption{
{\bf Effect of Noise and Biases from the Embedding Networks.}
\textbullet$\;${\bf(a)} Information flux in the core, measured by a flux indicator, in five different configurations: Core receiving the full signals from the embedding networks; Core receiving only temporal averages (biases); Biases plus normal distributed noise; Uniform ($b_i\is b$) optimal biases; Individually optimized biases.
\textbullet$\;${\bf(b)} Flux indicator as a function of a uniform bias, applied to all core neurons.
\textbullet$\;${\bf(c)} Flux indicator as a function of the iteration step in an evolutionary optimization of the core neuron's individual biases.
\textbullet$\;${\bf(d)} Optimal individual bias values for the core neurons, found by the evolutionary procedure.
\textbullet$\;${\bf(e)} On-probabilities of the core neurons when the optimal biases are applied. 
\textbullet$\;${\bf(f)} Individual biases as a function of time, when each neuron independently regulates its time-averaged output signal to $1/2$ with an adaptive feedback loop. The dashed lines are the globally optimal bias values found with the evolutionary procedure.
}
\label{Fig_F}
\end{figure}

\NI In our next numerical experiment, we return to the isolated core and sweep the value of a uniform (identical) bias $b_i=b$ applied to all core neurons (Fig.\ref{Fig_F}(b)). We find that the flux indicator peaks at $b\is-7.5$, where it reaches a value of $f\is187.4$, significantly exceeding even the reference value of $f\is139.3$ obtained when the core is driven by the full signals from the embedding network.

\NI Since individually different biases promise an even larger information flux in the core, we next perform an evolutionary optimization of the ten-dimensional bias vector $\vec{b}=(b_0,\ldots,b_9)$, using different initial conditions. As expected, the flux indicator increases monotonically as a function of the evolutionary iteration counter (Fig.\ref{Fig_F}(c)), irrespective of the starting vector (compare the four restarts). In particular, when starting from the uniform bias (blue curve), the flux indicator grows from $f\is187.5$ to clearly higher values. The globally best result of the evolutionary procedure (red curve) yields a flux indicator of $f\is203.9$.

\NI It turns out that, in the globally optimal case, all individual biases $b_i$ differ from each other, but are all negative (Fig.\ref{Fig_F}(d)).

\NI Remarkably, when these globally optimal biases are applied to the isolated core network, all ten neurons operate at a time-averaged on-probability of $p_1\is\frac{1}{2}$, the value that maximizes the output entropy of each individual neuron (Fig.\ref{Fig_F}(e)) and thereby places the network in a regime of high information flux.

\NI Finally, we test whether simple adaptive mechanisms, acting independently in each neuron, can also drive the system into this favorable maximum-entropy regime. For this purpose, we assume that each neuron adapts its individual bias such that its time-averaged output approaches $1/2$ (compare Methods \ref{MethAdaptBias}). Indeed, when all biases are initialized to zero, this mechanism converges to the globally optimal bias values found by the evolutionary flux-optimization procedure (Fig.\ref{Fig_F}(f)).

% *******************************************************
\section{Discussion}
% *******************************************************

\subsection{Summary}

\NI In this work, we implemented and analyzed a simplified toy model of layer 5 cortical microcircuits, in which a small core population of strongly connected neurons is embedded within larger peripheral and interneuron populations. The structure and connectivity parameters of the model were adopted, as far as possible, from experimental observations \cite{song2005highly}, while the remaining unconstrained parameters were adjusted such that the network operated in a dynamically useful regime.

\NI We found that the information flux in the embedded core is drastically larger than in the isolated core. In principle, this flux-enhancing function of the embedding network could arise from a feedback loop, in which the core sends signals to the embedding network, which in turn sends control signals back into the core. However, simulated lesion experiments demonstrated that signals from the core to the embedding network are not required for its flux-enhancing effect.

\NI The remaining question was whether the control signals from the embedding network to the core are beneficial mainly because of their temporal averages, acting as effective biases, or because of their random fluctuations. We found that both contributions are significant, but that the biases alone already raise the information flux of the core well above the level of the isolated core. Interestingly, adding independent Gaussian white noise to these biases decreases the information flux, indicating that the specific non-Gaussian statistics of the embedding-network control signals is essential.

\NI In a further numerical experiment, we evolutionarily optimized the 10-dimensional vector of individual biases for the core neurons and found that applying these optimal biases yields the largest information flux observed throughout this study. Under this condition, all core neurons attain a time-averaged on-probability of $1/2$, corresponding to maximal individual output entropy.

\NI Finally, we demonstrated that a simple self-organizing mechanism, acting locally within each core neuron, can automatically find these optimal, information-flux-maximizing biases and can also restore the system to this optimal regime after a perturbation.

% *******************************************************

\subsection{Relevance of the Results}

\NI Our results suggest that the embedded-core structure of layer 5 cortical microcircuits indeed ensures a large information flux in the core, brought about by a combination of effective biases and non-Gaussian random fluctuations. We did not find evidence, however, that the overall system operates as a closed feedback loop.

\NI In technical applications of recurrent neural networks (RNNs), such as reservoir computing systems, our findings suggest that it is important to place the information-processing network into a suitable operating regime with large individual output entropy of the neurons. Moreover, we have shown that such a regime can be established automatically through simple adaptive mechanisms acting locally within each neuron.

\NI From a more methodological perspective, we have demonstrated how the numerical implementation of biological networks, even if only approximate, opens these highly evolved systems to detailed reverse-engineering studies that would not be feasible in the biological system itself. Given the increasingly sophisticated methods for extracting structural and dynamical parameters from biological systems, this approach could become an important new branch of theoretical biology in the future.

% *******************************************************

\subsection{Open Questions and Outlook}

\NI We began the investigation of our model system with the natural assumption that the strongly connected core of neurons is primarily responsible for the actual information processing, while the more weakly connected background populations of excitatory and inhibitory neurons mainly serve a supporting role for the processing activity in the core. Future work should examine, however, whether the numerous weak connections in the embedding network may themselves contribute more actively to information processing.

\NI Moreover, our second working assumption -- namely that efficient information processing requires a large information flux within the network -- also deserves a more rigorous investigation in the context of different computational tasks. On the one hand, it is clear that a network with zero information flux, that is, zero mutual information between successive network states, lacks any degree of determinism and is therefore useful only for the generation of random signals. On the other hand, a network with maximal information flux would evolve with perfect determinism along a cyclic attractor, without stochastic deviations from its trajectory. In such a regime, tasks requiring probabilistic or creative outputs would no longer be possible. An interesting future research direction will therefore be to continuously tune the information flux of a network through suitable control parameters while simultaneously measuring its performance across different computational tasks.

\NI Since we found that adding simple Gaussian noise to a network with already suitable neural biases reduces the information flux, whereas the non-Gaussian signals originating from the embedding network increase it, our results also motivate a more systematic study of the recurrence resonance phenomenon under complex, non-Gaussian, and possibly temporally correlated forms of noise. 

% *******************************************************

\subsection{Relation to other Works}

\NI The present work connects to several distinct lines of research in computational neuroscience, recurrent network theory, and dynamical systems.

\NI The finding that information flux is maximized when core neurons operate near a high-entropy, balanced firing regime resonates with the well-established "edge of chaos" concept in reservoir computing \cite{langton1990computation}. A substantial body of work has shown that computational capacity in recurrent networks peaks near the boundary between ordered (fixed-point or oscillatory) and chaotic dynamics \cite{bertschinger2004real}, and that task performance is often maximized in this transitional regime [30]. Our study extends this line of thinking by showing that maximizing individual neuron entropy — achievable through neuron-specific bias tuning — is a principled and local route to reaching this favorable operating point. Compared to global parameter tuning such as scaling the spectral radius of the weight matrix, our approach has the advantage of being implementable through biologically plausible, locally acting mechanisms.

\NI The Recurrence Resonance mechanism that we build upon [19, 39] is closely related to the classical stochastic resonance phenomenon, in which an optimal amount of noise enhances signal detection or processing in a nonlinear system \cite{benzi1981mechanism,wiesenfeld1995stochastic}. Stochastic resonance has been implicated in a range of biological sensory processes [36, 37, 38] and has also been leveraged deliberately to improve performance in artificial neural systems [35]. Our results extend the standard stochastic resonance picture in an important direction: the embedding network does not inject simple Gaussian white noise, but rather structured, non-Gaussian signals that outperform Gaussian noise of matched variance. This finding motivates a more systematic study of recurrence resonance under complex, temporally structured, or non-Gaussian noise, a direction that has received comparatively little attention in the existing literature.

\NI The lesion experiments reported in section 3.5 reveal that the flux-enhancing influence of the embedding network operates in a strictly feedforward manner: removing outgoing projections from the core to the surrounding populations leaves the information flux in the core essentially unchanged. This is reminiscent of the classical distinction between driving and modulatory inputs in cortical circuits \cite{sherman1998actions}, in which subcortical or feedback projections modulate the gain or operating point of a target area without transmitting detailed content. Our finding that the peripheral and interneuron populations effectively set the bias of core neurons — rather than closing a regulatory feedback loop — aligns with this interpretation. It is also noteworthy in the context of predictive coding frameworks \cite{rao1999predictive,friston2005theory}, where top-down signals are thought to convey predictions that shape the dynamic regime of lower-level areas, rather than content per se.

\NI The adaptive bias rule introduced in section 2.4 is functionally equivalent to a homeostatic plasticity mechanism: each neuron independently adjusts its excitability to maintain a target firing rate of one half. This is closely related to intrinsic excitability regulation observed in biological neurons \cite{turrigiano2004homeostatic}, as well as to theoretical proposals such as the Bienenstock-Cooper-Munro (BCM) learning rule \cite{bienenstock1982theory} and firing-rate homeostasis models \cite{turrigiano2008self}, in which neurons regulate their own gain to prevent runaway excitation or silence. What our analysis adds to this picture is a precise information-theoretic justification for the target firing rate: a time-averaged on-probability of one half maximizes the individual output entropy of a binary neuron, which in turn maximizes the attainable information flux. This links homeostatic regulation directly to information-theoretic optimality, a connection that deserves further investigation in both biological and artificial systems.

\NI The embedded-core architecture that forms the basis of our model — a densely and strongly coupled hub embedded within a sparser background network — is an instance of the core-periphery organization that has been widely documented in complex networks, from the World Wide Web to protein interaction networks \cite{colizza2006detecting}. In neural systems, this motif appears at multiple scales. At the level of large-scale brain connectivity, it is closely related to the "rich club" organization identified in human white-matter connectomes \cite{van2011rich}, where a small set of highly connected hub regions are preferentially interconnected with one another. At the local circuit level, the layer 5 microcircuit data of Song et al. [23], on which our model is based, exemplifies the same principle: a small fraction of neuron pairs share exceptionally strong synaptic connections that form a structural backbone. Our results suggest a functional interpretation of this ubiquitous architectural motif: the core-periphery structure may be shaped, at least in part, by selective pressure to maximize information flux within the core, with the peripheral network providing the biasing and fluctuation signals necessary to sustain a high-entropy dynamical regime. Whether a similar interpretation applies at the large-scale connectome level remains an open and intriguing question.

% *******************************************************
\section{Additional Information}
% *******************************************************

\subsection{Author contributions}

CM supervised the study and wrote the paper. AG and KP implemented the methods and evaluated the data. AS discussed the results and acquired funding. AM and TK discussed the results and provided resources. PK conceived the study, discussed the results, acquired funding and wrote the paper.

\subsection{Funding}
This work was funded by the Deutsche Forschungsgemeinschaft (DFG, German Research Foundation): grants KR\,5148/3-1 (project number 510395418), KR\,5148/5-1 (project number 542747151), KR\,5148/10-1 (project number 563909707) and GRK\,2839 (project number 468527017) to PK, and grants SCHI\,1482/3-1 (project number 451810794) and SCHI\,1482/6-1 (project number 563909707) to AS.

\subsection{Competing interests statement}
The authors declare no competing interests.

\subsection{Data availability statement} 
The complete data and analysis programs will be made available upon reasonable request.

\subsection{Third party rights}
All material used in the paper are the intellectual property of the authors.

% ***************************************************
% REFERENCES
% ***************************************************

\newpage
\bibliographystyle{unsrt}
\bibliography{references}

% ***************************************************
% FIGURES
% ***************************************************

% -------------------------------------------------
\newpage
% -------------------------------------------------

% -------------------------------------------------
\newpage
% -------------------------------------------------

% -------------------------------------------------
\newpage
% -------------------------------------------------

% -------------------------------------------------
\newpage
% -------------------------------------------------

% -------------------------------------------------
\newpage
% -------------------------------------------------

% -------------------------------------------------
\newpage
% -------------------------------------------------

\end{document}